# The Dimensions of Data Labor: A Road Map for Researchers, Activists, and Policymakers to Empower Data Producers

Hanlin Li
hanlinl@berkeley.edu
University of California, Berkeley
Berkeley, CA, USA

Nicholas Vincent
nickvincent@u.northwestern.edu
University of California, Davis
Davis, CA, USA

Stevie Chancellor
steviec@umn.edu
University of Minnesota
Minneapolis, MN, USA

Brent Hecht
bhecht@northwestern.edu
Northwestern University
Evanston, IL, USA

## ABSTRACT

Many recent technological advances (e.g. ChatGPT and search engines) are possible only because of massive amounts of user-generated data produced through user interactions with computing systems or scraped from the web (e.g. behavior logs, user-generated content, and artwork). However, data producers have little say in what data is captured, how it is used, or who it benefits. Organizations with the ability to access and process this data, e.g. OpenAI and Google, possess immense power in shaping the technology landscape. By synthesizing related literature that reconceptualizes the production of data for computing as "data labor", we outline opportunities for researchers, policymakers, and activists to empower data producers in their relationship with tech companies, e.g advocating for transparency about data reuse, creating feedback channels between data producers and companies, and potentially developing mechanisms to share data's revenue more broadly. In doing so, we characterize data labor with six important dimensions - legibility, end-use awareness, collaboration requirement, openness, replaceability, and livelihood overlap - based on the parallels between data labor and various other types of labor in the computing literature.

## 1 INTRODUCTION

Technology users generate large troves of data in their daily interactions with computing systems, e.g. behavior logs, content, and personal information. Currently, this data primarily benefits just a small set of technology organizations that are equipped with the means and resources to collect, process, and model data at scale for their own benefits (e.g. insights, models, sales of services and advertisements). For example, publicly available texts and artwork enabled the creation of generative AI models like ChatGPT and Dall-E because model developers were able to scrape and process data from billions of web pages [1]. Conversely, data producers like artists, writers, and users have little to no power in deciding how their data is used or who it benefits [4, 7, 42, 63]. This power imbalance between data producers and technology operators has manifested in public outcries about industry practices in the tech sector. For example, emerging generative AI models such as Stable Diffusion, Dall-E, and GitHub Copilot have sparked extensive criticism among artists and programmers because of these models' unapproved reuse of their work and implications on future employment opportunities [79–81]. More broadly, social media users have long protested the monetization of user data and the corporate surveillance practices that tend to go with it [45].

Given data producers' lack of power over the data they generate, researchers, policymakers, and activists have advocated for a new producer-oriented paradigm shift to increase the voice of the data-generating public – understanding data generation as a form of labor, or "data labor" [7]. Supporters of this approach have argued that treating data as an outcome of social labor instead of "exhaust" will pave the way for more broadly distributing the power and benefits of data [86], and scholars have addressed what this may look like in practice. Initial (yet abstract) proposals include supporting "data unions" [63] or "mediators of individual data" [42] that negotiate data use terms with technology firms on behalf of their data-producing "union" members [63], drafting legislation that would grant users greater control over the data they produce [1, 76], and creating tools to support user-driven collective action [20, 86].

---
[1]https://commoncrawl.org/2022/10/sep-oct-2022-crawl-archive-now-available/

Despite these abstract "blueprints" for reconceptualizing data generation as labor, the research, policy, and advocacy community is missing a transformation of these initial ideas into more concrete and actionable guidelines [78]. Given that data producers contribute to data-driven technologies in a myriad of ways, an explicit characterization of data labor is crucial to guide researchers, data producers, and policymakers to concretely address the power imbalance between the public and large technology firms. More specifically, such characterization will accentuate how different types of data labor may require different strategies and interventions to empower data producers through research, development, and policy practices.

This paper provides an *actionable road map* for researchers, activists, and policymakers to empower data producers by synthesizing literature from different relevant disciplines into six key dimensions of data labor: *legibility, end-use awareness, collaboration requirement, openness, replaceability, and livelihood overlap*. While the prevalent interpretation of *labor* implies compensation, our definition of data labor encompasses both compensated data production (e.g. labeling images on Amazon Mechanical Turk for a computer vision company) and uncompensated data production (e.g. generating behavior logs). Our road map is informed by complementary frameworks of empowerment introduced by Schneider et al. [72] and data leverage from Vincent et al. [86].

The six dimensions of data labor we detail in this work are not intended to be all-encompassing; rather, they aim to serve as an important step towards having more action-oriented conversations about data labor and advancing scholarly discussion on the relationship between data, power, and social inequality [14, 22, 27]. We also discuss opportunities for future work to identify potential dimensions.

## 2 DEFINITION AND RELATED WORK

### 2.1 Defining Data Labor

Discussions of "data labor" have not yet coalesced on a concrete definition and largely operate at a conceptual level. In 2018, Arrieta-Ibarra et al. asked "should we treat data as labor?", in response to the lack of recognition of users' role in the advancement of technology and the data economy [7]. Recently, this approach has motivated economists, legal scholars, and computing researchers to explore potential implications of treating user activities as labor via simulations [10]. For example, Jones and Tonetti simulated how granting users rights to the data they produce and allowing the data to be used across firms can maximize social gains from the data economy [39]. Others have taken a step further and recommended establishing third-party intermediaries that are analogous to labor unions to facilitate the relations between subgroups of users and technology companies [10, 42]. In a similar vein, practitioners have piloted applications and platforms that allow users to control who has access to the data they generate (e.g. the Solid project[2] and Streamr[3]).

For discussions of "data labor" to have meaningful, pragmatic implications for data producers, we first needed to distinguish what is and is not "data labor". Drawing from emergent discussions about

[2]https://solidproject.org/
[3]https://streamr.network/

data labor across disciplines [7, 39, 66], we offer a working definition:

> *Activities that produce digital records useful for capital generation.*

Said differently, an activity must meet two criteria to be labeled as data labor: 1) it creates or enhances data ("digital records"), and 2) the resulting data helps to generate capital. Correspondingly, those who contribute data labor have a role as "data laborers".

In this paper, we specifically focus on data labor that subsidizes prominent tech companies because of the emerging, substantial power inequity between these entities and the data-generating public. While governmental agencies (e.g. census bureaus), research organizations, civil societies, and non-profit organizations (e.g. the Wikimedia Foundation) also rely on data labor to generate capital (see Discussion for details), they do so to a lesser extent than prominent tech companies.

With data playing a crucial role in driving tech innovation (e.g. [3, 11, 85]), we foresee that more data generation activities will fall under this definition of data labor. As we will discuss much more below, it is possible (and indeed very common) that under this definition, many people are performing data labor without being aware of its additional value or potential for revenue generation, such as posting articles that are later useful for training large language models. Indeed, the recent rise in popularity of tools like ChatGPT that rely on massive scraping has shown that many content production activities on the web may fall into the scope of data labor.

Beyond Arrieta-Ibarra et al.'s work [7], many other research areas and scholarships have explored concepts related to data labor. Below we relate our notion of data labor to similar concepts of labor and work in literature.

*2.1.1 From Computer-Mediated Labor to "Data Labor".*
Computer-mediated labor has been the focus of human-computer interaction (HCI) and computer-supported cooperative work (CSCW) research and we briefly examined the activities that these scholars consider to be labor to inform our definition of data labor.

Computer-mediated labor studied in HCI and CSCW encompasses both compensated and uncompensated activities, and so does our definition of data labor. While compensated work is a common type of labor that HCI and CSCW researchers study, including desk work, gig work (driving for Uber/Lyft), crowdwork (Amazon Mechanical Turk tasks), and low-wage work [25], the scope of labor also includes "settings in everyday life", ranging from domestic labor, to leisure activities, to social networking [18].

In a similar vein, we define data labor as encompassing both witting labor activities such labeling images on Mechanical Turk and unwitting ones such as producing content on a social network. Moreover, there exists a substantial overlap between computer-mediated labor and data labor: crowdsourcing [70], peer production [29], and content moderation [26] are all data labor that advances computing systems and, thereby, benefits technology companies financially.

### 2.1.2 Digital Labor.
*Digital labor* is another term that has been used to refer to monetized online activities, regardless of whether they occur at traditional workplaces or are compensated [73]. In particular, Terranova argued that "the Internet is animated by cultural and technical labor through and through, a continuous production of value that is completely immanent to the flows of the network society at large" [75]. This work has subsequently inspired in-depth examinations of how online interactions generated value with the commercialization of the internet, e.g. interacting with YouTube videos [64] and managing communities [49].

While in most cases, prior work on digital labor is relevant to data labor as we've defined here, there may be some examples of digital labor that are not data labor. For instance, private communication activities like participating in listservs do not necessarily result in capital creation. Furthermore, not all types of data labor are likely to be considered digital labor. Passively produced data such as location data, traffic patterns, and private preference information actively play a role in the improvement of advertising models, navigation algorithms, and commercial recommender systems – however, they are not commonly seen as digital labor [73].

### 2.1.3 Crowdwork.
*Crowdwork* is a subcategory of computer-mediated labor in which crowds of distributed laborers complete small-scale tasks for payment. Examples include completing image labeling tasks that enabled the creation of ImageNet [21] and producing texts used to train spam filters [58]. Many instances of crowdwork are unambiguously data labor, although there may be exceptions: completing behavior experiments on Amazon Mechanical Turk run by academic institutes–a prominent type of crowdwork [33]–does not always lead to capital generation, and therefore, may not be data labor.

### 2.1.4 Data Work in Data Science and Machine Learning.
*Data work* is a relatively new term that emerged from calls for shifting attention to data that powers intelligent systems. The term has been used to loosely encompasses both data generation by users and data labeling and cleaning by crowdworkers and technologists such as data scientists [23, 53, 69]. For example, Sambasivan et al. referred to data work as the activities by those who are "data science workers" (a role discussed at length by Zhang et al. [87]). In another example, Møller et al. studied hospital workers and referred their production of quality data about patient records as data work [53]. Finally, Miceli and Posada defined data work as define as "the labor involved in the collection, curation, classification, labeling, and verification of data" [51].

Data labor is a slice of the broadly-defined data work [51]. It concerns primarily the upstream activity of generation, not downstream processing activities such as data cleaning and filtering. It is certainly possible that data scientists who clean up datasets also perform data labor, e.g. labeling images for their own models.

## 2.2 Frameworks of Power and Data Leverage

This work serves as an extension of prior literature that highlighted the importance of examining the power imbalance and social inequalities in computing systems [14, 27, 52]. For example, Ekbia et al. called for researchers and designers to take into account the broader political economy of computing and pursue a more equitable distribution of wealth as the ultimate goal [27]. In response to such calls, we constructed our roadmap to provide some concrete pathways toward more equitable social relationships. Below, we describe how this roadmap is built upon the concept of power and data leverage from prior work.

To empower data labor, it is critical to adopt an appropriate definition of power, and what it means to shift more power to data producers. The concept of "power" is hotly debated [34]. We draw on Schneider et al.'s work on HCI and empowerment that is intended to "add structure and terminological clarity" to the notion of empowerment in computing research [72]. Through a synthesis of studies on empowerment, Schneider et al. highlight the distinction between two notions of power: power-to and power-over [72]. Power-to corresponds to an individual's "ability to do something" (Schneider et al. drew on Arendt's work for this definition [6]). Applying this notion to data labor, power-to means that people can freely make decisions around their data labor, i.e. choosing not to participate in activities that are data-generating or deleting the data resulting from this labor. Power-over refers to "the relation between multiple actors" [72] or in Dahl's words, "A has power over B to the extent that he can get B to do something that B would not otherwise do" [19]. In the context of data labor, power-over means that people can influence those who currently benefit from their labor – technology operators – around decisions regarding data-driven technology.

Vincent et al.'s framework of *data leverage* further highlights connections between power-to and power-over [86]. Data leverage describes how data producers may influence technology operators through three "levers": "data strikes" and "data poisoning" harm a technology operator, while "conscious data contribution" can boost up an alternative operator [86]. Each data lever requires collective action to be effective in which individuals engage via activities like withholding data or manipulating data as a group. Performing these specific actions requires "power-to" (e.g. a legal or technical guarantee that users can delete their data contributions). Power-over is achieved when a critical mass [57] of participation is reached.

## 3 THE DIMENSIONS OF DATA LABOR

We describe six key dimensions of data labor: *legibility, end-use awareness, collaboration requirement, openness, replaceability, and livelihood overlap*. While each dimension is a spectrum, we provide examples of data labor that fall on the ends of each spectrum to show how even a relatively dichotomous understanding of each dimension can provide immediately usable insights. For each dimension, we provide 1) an assessment of how data labor's position along the dimension is related to power and 2) potential opportunities to empower data laborers who are at different positions along the dimension.

## 3.1 Process of Dimension Construction

Our research team collectively synthesized dimensions that can characterize data labor. We drew from discussion around data labor (e.g. [7, 67]) and its related concepts mentioned above (e.g.

computer-mediated labor and crowdwork) to identify potential dimensions that may be useful to characterize these data-generating activities. We then discussed what dimensions may be correlated and therefore should be merged. Our research team then ranked and chose the dimensions that can most clearly delineate the wide range of data labor activities happening in real life through an iterative process.

As mentioned above, these dimensions are not meant to be exhaustive and given the ever-evolving landscape of data-driven technologies, we expect more dimensions to emerge as the public's data labor comes in new forms or is used in novel ways by technology operators.

## 3.2 Legibility: Do Data Laborers Know Their Labor Is Being Captured?

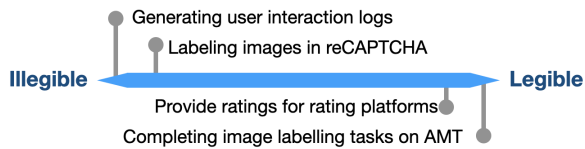

**Figure 1: The dimension of legibility: do data laborers know their labor is being captured? As in all figures in this section, the positioning of data labor instances on the spectrum is approximate and may vary depending on the specific case. For example, for those who understand the background of reCAPTCHA, labeling its images may be legible data labor.**

Data labor can be *illegible* or *legible*, depending on whether data laborers know their labor is being captured (Fig. 1). When a person's activity leads to digital records being created or enhanced without the person's knowledge, this is illegible data labor.

Examples of illegible data labor include the generation of user interaction logs (for search engines, ad systems, recommender systems) and the production of image labels when users complete reCAPTCHA. It is worth noting that legibility depends on a person's knowledge about the system they use: two different people using the system can produce the same output with different levels of knowledge about the capture of their data labor.

On the other end of the legibility spectrum exists *legible* data labor – activities that people can clearly see being captured by technology companies. Examples include generating ratings for rating platforms such as Yelp and Google Maps and completing a crowdwork image labeling task in which the collection of labels is clearly disclosed.

*3.2.1 Relationship with Power and Its Implications for Data Labor.*
Legibility is a necessity for data laborers to exert their power over technology companies. When people do not realize that they are performing data labor, this naturally inhibits their power to withhold or change this labor. This illegibility further limits opportunities for collective action (e.g. a data strike [84]) or other types of action (e.g. calling on regulators).

**Empowering illegible data labor:** Mitigating illegibility – i.e., moving illegible data labor to the legible end of the spectrum – is a first step towards empowering illegible data labor. Given that making labor's value transparent helps workers' collective negotiation with employers in traditional labor advocacy [40], it is likely effective to apply this tactic to illegible data labor. Researchers and activists may develop tools that measure and communicate the economic or utility value of data labor, such as the Facebook Data Valuation Tool (FDVT), a tool that calculates the worth of Facebook users' attention in real time [31]. By making data legible, these tools can equip data laborers with knowledge about how to effectively leverage their data labor against technology companies.

Activists can explore how existing technologies that disrupt the collection of data can help to make illegible data labor more legible. Currently, privacy-preserving technologies such as anti-tracking browser extensions and protest-assisting technologies such as ad blockers are actively preventing data labor from being used by tech companies and have gained a considerable user base [46]. Activists may experiment with providing add-on features to these technologies to highlight the "cost" or "lost ad revenue" users have caused to highlight the value of data labor (e.g. [43]).

It is worth noting that making illegible data labor legible does not always translate to power for data producers. They may not always have power-to, i.e. control over their data labor due to external social constraints. For instance, when faced with suspicious, privacy-violating requests, crowdworkers may still complete them because of their need for extra income [71].

**Empowering legible data labor:** In contrast, people performing legible data labor are most likely equipped with more power-to than the illegible condition. They may simply choose not to volunteer or complete a task, so they do not produce any data labor as seen in various non-use cases [8, 45]. There is also early evidence about those that perform legible data labor exerting their power over technology operators through collective action. For example, Reddit volunteer moderators collectively negotiated with Reddit for better moderation tools [47] and crowdworkers leveraged Turkopticon to improve their working conditions on Amazon Mechanical Turk [37].

Those performing legible data labor can immediately benefit from research and tools that strengthen their power over companies through collective action. Past research has shown through observation [48] and through simulation [84] that if users collectively withhold their data labor, i.e. data strikes, they can negatively affect data labor-dependent platforms. Given this potential, researchers may want to further study how to help data producers organize collective action to influence technology operators, answering questions about how different conditions and techniques affect data producers' participation and how to create revenue loss or performance loss for technology operators.

For collective action to succeed, it will be important to ensure that individual data producers have meaningful control over their data output. Data regulations like GDPR (General Data Protection Regulation) and CCPA (California Consumer Privacy Act) [55] provide legal ground and infrastructure for data producers to exercise their right to deletion and right to portability. Broadening the types of data covered by such regulations, i.e. from personal data to other types of data such as behavior logs and crowdsourced datasets of image labels will help to further pave the way for data producers to collectively leverage their data.

## 3.3 End-Use Awareness: Do Data Laborers Understand How Labor Is Used to Generate Capital?

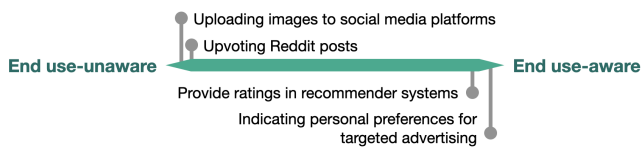

Figure 2: The dimension of end-use awareness: do data laborers understand how labor is used to generate capital? The positioning of data labor instances on the spectrum is approximate and may vary depending on the specific case.

While legibility characterizes whether data producers have the knowledge of their labor being *captured*, end-use awareness characterizes the degree to which they are aware of how the resulting data is *used* downstream to generate capital (Fig. 2). One can have full knowledge of their labor being captured, or put another way, perform fully legible data labor, but have no awareness of their labor's impact. Currently, a variety of data-driven technology companies benefit from data labor without the laborers' knowledge. For example, content creators such as Wikipedia editors and journalists are well aware of their data labor being captured by online platforms, but may not be fully aware of how this labor is being used downstream, e.g. for search engine performance improvement [50, 83] and training large language models [65].

End-use awareness may change drastically over time as technology operators keep identifying innovative ways to utilize and monetize data labor. In other words, end-use awareness is naturally dependent on what end-use cases exist. Recent advances in generative AI models (e.g. GPT-4 [11] and MusicLM [3]) have shown that much of the legible data labor people performed before those advances, from writing articles to publishing music online, now fall under the end use-unaware end of the spectrum. Moreover, similarly to legibility, end-use awareness may vary from person to person; for example, for the act of uploading images to social media platforms, a computer vision researcher has very likely more end-use awareness than an average user does.

End use-aware data labor occurs in scenarios in which data producers are informed of or have some understanding of how the output of their labor is being used. Examples include targeted advertising, personalized newsfeed algorithms, and recommender systems. Through direct interactions with these technologies' interfaces, data producers learn that their actions are being used to support certain functions or features of these systems.

*3.3.1 Relationship with Power and Its Implications for Data Labor.* In general, end-use awareness is likely to lead to more power to data labor. If one understands the downstream capital generation implications of the data they produced, it is possible to alter such labor purposefully. For example, users reported that they would be demotivated if their contribution to Wikidata, a structured, public access database analogous to Wikipedia, were to be used primarily for profits [88].

**Empowering end use-unaware data labor:** While it is understandable that technology operators have incentives to keep their data-dependent technologies proprietary and reveal few details (i.e. what data they use, and for what purposes), this practice tends to reduce end-use awareness and, therefore, disempowers the public. Thus, a key direction for researchers and activists is to increase end-use awareness where possible. For instance, such a tool might tell a Wikipedia contributor that their edits on a particular article appeared in Google's knowledge panel or are being used by certain large language models.

Policymakers can also play a role by mandating end-use awareness, particularly for sensitive data such as biometric data and personal information. Future policies on data use may focus on requiring technology companies to disclose how such data will be used downstream. Moreover, mandating a mechanism for opting out of model training (e.g. Spawning AI's 'haveibeentrained.com', an opt-out form that has been honored by Stability AI) will provide data laborers individual control over the end use cases of their work. Of course, an ideal system might involve pushing more domains of data and content towards an opt-in paradigm, though this may be an uphill battle and require opt-out as an intermediate step.

Relatedly, understanding how moving end use-unaware data labor to the other end of the spectrum may affect data labor-dependent technologies is also a fruitful area for research. Data producers' concerns about the end use of data labor may disincentivize the production of data labor. As such, increases in end-use awareness may inadvertently reduce the utility of technologies that are currently providing enormous benefits to the public, e.g. Wikipedia [50, 85], mental health communities [15], and review platforms [44], a potential risk that warrants further investigation.

**Empowering end use-aware data labor:** For end use-aware data labor, given that there already exists significant public use awareness of certain types of data labor in targeted advertising and social media[13], researchers and activists may be interested in focusing on these types of data labor and investigating how to transform current end-use awareness to collective action. For example, activists may consider developing tools that make straightforward how use of data may be affected negatively by data producers collectively withholding or poisoning data. Specifically, such tools may draw from proof-of-concept studies in HCI such as AdNausem [36] and Out of Site [43] that illustrate the downstream effects of users' protesting actions such as data strikes or data poisoning (e.g. "deleting your data would incur loss of ad sales to Facebook").

Researchers and policymakers may also consider strengthening existing legal frameworks that allow users to proactively control how data is being used downstream (in addition to opting out mechanisms as mentioned above). For example, users may be able to attach licensing clauses such as "Do not use for surveilliance" in addition to "Do not share with third parties" to the data they produce [17].

## 3.4 Collaboration Requirement: Does Data Labor Involve Collaboration?

Data labor activities can be mapped onto a spectrum from non-collaborative to collaborative based on the extent to which data laborers work together (Fig. 3). This dimension is informed by the distinction between team work and individual work in computer-mediated labor.

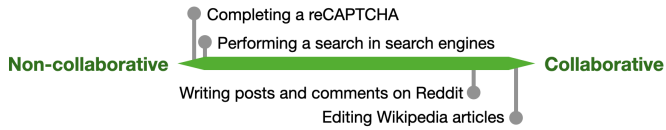

Figure 3: The dimension of collaboration requirement: does data labor involve collaboration? The positioning of data labor instances on the spectrum is approximate and may vary depending on the specific case.

Non-collaborative data labor activities are those that data producers perform in isolation, such as completing a reCAPTCHA, completing data labeling tasks on platforms like Amazon Mechanical Turk, using a search engine or personal recommender system.

When users' data-generating activities involve elements of deliberation, communication, and other forms of teamwork, this is collaborative data labor. This type of data labor is prevalent in social computing systems in which data laborers actively communicate with each other and make decisions about the division of labor themselves. For example, writing posts and comments, a prominent form of data labor that benefits Reddit and operators of language models such as GPT-4 [11], is largely driven by data producers' intrinsic motivation to produce content and interact with other community members.

*3.4.1 Relationship with Power and Its Implications for Data Labor.* The relation between power and collaboration requirement is complicated by the fact that social connections between data laborers can facilitate collective action but also creates costs for those withholding data labor. Data laborers who perform non-collaborative data labor can easily stop performing their tasks without worrying about losing connections with their peers or endangering a collaborative project. However, because of the lack of collaboration in this data labor, users lack shared identity, making it difficult to organize collectively and gain power to influence technology operators. Conversely, those who perform collaborative data labor can theoretically leverage their network for collective action against their "employers", i.e. technology companies. For example, historically, Reddit users coordinated their exit from the platform due to disagreements with the platform's changing policies [54]. However, they may face social cost to exert power-to by withholding or changing their labor if those in their close network are not doing so simultaneously. For example, a Wikipedia editor who does not want their data labor being exploited by for-profit companies and contemplates leaving Wikipedia may fear losing connections with their community members.

**Empowering non-collaborative data labor:** One step towards empowering non-collaborative data labor is to create connections between users for collective action. In traditional labor organizing and crowdworker organizing, workers benefited from having a shared professional identity to pave the way for collective action [32]. As such, to empower data labor, researchers and activists may explore when it is possible to foster a sense of community among users who perform non-collaborative data labor such as Amazon product reviewers. Moreover, activists may benefit from prior HCI research on overcoming challenges associated with collective action by a dispersed, or very loosely connected labor force. For example, by studying crowdworkers' collective labor advocacy efforts, Salehi and colleagues identified two key issues–losing momentum and community frictions–and made corresponding suggestions for design to mitigate these issues [68]. This prior work can serve as an exemplar to inform future research on how non-collaborative data labor can be effectively organized to advance data producers' shared goals.

**Empowering collaborative data labor:** For collaborative data labor, researchers and activists may explore how to grant data producers greater control over their labor while still contributing to their teams and communities. This may be achieved by building alternative technologies that allow users to migrate but stay connected with their former community in some way. Prior work by Fiesler and Dym on fandom communities' platform migration have provided some concrete guidance on this direction [28]. Specifically, they have recommended that alternative technologies allow cross-posting and support data import by users who have had extensive history and interactions on their previous technologies. Policymakers may also be able to play a meaningful role in empowering data labor by mandating or otherwise supporting data portability so data laborers could more easily travel across technologies with their communities.

### 3.5 Openness: Is the Data Resulting from Data Labor Open for Use?

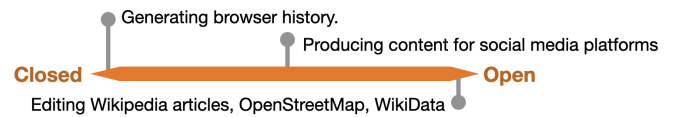

Figure 4: The dimension of openness: is the data resulting from data labor open for use? The positioning of data labor instances on the spectrum is approximate and may vary depending on the specific case.

Data labor's *openness* is characterized by how accessible the downstream data is to the public (Fig. 4). This dimension is informed by the prominence of the computer-mediated labor that supports open source projects and platforms [30]. Data labor captured by private systems to benefit specific individuals or groups and excludes others is closed, whereas data labor in systems that adopt copyleft licenses and make the fruits of data labor public is considered maximally open (e.g. Wikipedia[50], OpenStreetMap [5], and WikiData). Generally, openness of data labor is determined by the sociotechnical systems in which labor occurs, though data scraping has enabled companies to collect data from sources that are semi-public or private such as profile pictures, product reviews, and artwork [82]. Moreover, policymakers may make data generated in some closed systems open, such as ride-hailing records [56].

Examples of closed data labor include producing browser logs, querying information in search engines, and generating location traces on smartphones. In each of these cases, the resulting data output is usually guarded carefully by technology operators.

Examples of open data labor can be widely seen in academic computing research. For example, the Pushshift Reddit dataset, which provides open access to Reddit content, was used by researchers to create automated moderation tools and language models [9]. Other prominent examples of datasets created by open data labor include the MovieLens dataset [35], Wikipedia articles, ImageNet [21], OpenStreetMap datasets [5], and U.S. census records. Moreover, the debut of generative AI models shows that publishing any content on the web may become a de facto instance of open data labor as this content will be accessible to AI firms via data scraping, whether the producer consents to it or not.

*3.5.1 Relationship with Power and Its Implications for Data Labor.* It is possible to gain power-to for closed data labor when data deletion requests are honored. For example, once users have permanently deactivated their Facebook accounts or deleted ratings they have published on review platforms, technology operators will lose at least part of the value from such data labor. Regulations such as GDPR in the European Union [1], the CCPA in California [55], and a variety of similar initiatives have laid the groundwork for the public to exercise their power-to, e.g.removing data from technology operators' records.

Moving to the other end of the spectrum, open data labor, by design, often has pro-social goals such as making knowledge accessible to all and incentivizing innovations; however, this openness makes it difficult for data laborers to control who can benefit from their work. Once the aggregated datasets are made publicly available for download, those who produced this open data labor have little power to exclude technology companies from benefiting from their labor. Even in situations in which users can request to delete their data (e.g. the Pushshift Reddit API), their requests are unlikely to affect all downloaded copies of open datasets that are being re-purposed by private technology companies. Even when a whole dataset is retracted, technology operators, practitioners, and researchers may still have access to and use its copies, without needing permission or seeking input from those involved with the creation of the dataset (see [59] for an overview).

**Empowering closed data labor:** Researchers have proposed new, conceptual data governance models to mitigate the tension between openness and control in data, that is, how to make data accessible without sacrificing data producers' control. The idea of data cooperatives [60] is one of them, aimed at organizing data laborers to collectively make decisions about data usage. With a focus on preserving sensitive information, this approach does not require making the outcome of closed data labor completely public, and, therefore, preserves data laborers' power-to if they wish to control how data is being captured and used downstream.

Additionally, policymakers may establish other channels through which data producers can collectively and democratically shape the future of their labor. Similar to how shareholder meetings are required for publicly traded companies, policymakers may mandate operators of data-driven technologies to have public channels of communication with data laborers to solicit their input and feedback.

**Empowering open data labor:** To empower open data labor, we first need a more comprehensive understanding of the myriad ways open datasets power private technologies. Gaining this knowledge will help to identify which open datasets underpin today's digital infrastructure and therefore inform activists and policymakers of what kinds of open data labor can be potentially leveraged. Several studies have laid the foundation for this research direction, focusing on Wikipedia datasets. Specifically, researchers have studied how Wikipedia data benefits Google Search [50, 83], Reddit, and StackOverflow [85], and commercial websites in general [62]. As Wikipedia datasets become commonly used in large language models, researchers may further investigate other economic and social benefits of Wikipedia editors' open data labor. Researchers may also broadly explore other prominent open datasets whose implications on the tech industry have not been extensively investigated, e.g. OpenStreetMap [5, 77]. Mapping out the pipelines through which open data labor powers technologies will help activists and policymakers understand when it may be possible for data producers to leverage their open data labor.

## 3.6 Replaceability: Is Data Labor Replaceable?

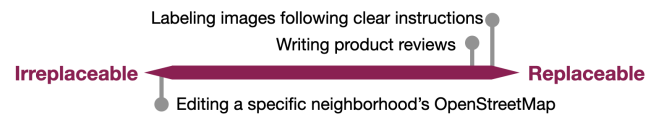

**Figure 5: The dimension of replaceability: Is data labor replaceable? The positioning of data labor instances on the spectrum is approximate and may vary depending on the specific case.**

Like computer-mediated labor, data labor may require certain background, knowledge, skills, or contextual aspects to perform, making it irreplaceable (Fig. 5). Replaceability can also be thought of similarly to the conceptions of skill in labor economics that emphasize modeling workers "endowments" to perform certain task [2]. For the majority of data-dependent systems that involve modeling the behavior and preferences of people, model performance will have some degree of specificity to the people who contributed data, i.e. performing irreplaceable data labor. A system for personalized advertising will require specific data about personal preferences, and a system that provides geographic knowledge will need content from a diverse set of regions that are irreplaceable by others [38].

There are many examples of data labor that are highly replaceable in the sense that many people, with a variety of backgrounds and contexts, could perform the data labor. Prominent examples include instruction-based labeling tasks, or simple tasks such as fixing typos in Wikipedia articles.

*3.6.1 Relationship with Power and Its Implications for Data Labor.* Data laborers responsible for irreplaceable data will have more ability to directly impact technology performance. This means a group of people capable of generating data that is currently underrepresented (e.g. people who can write in languages not currently captured in existing training data) may have power over those that benefit from their labor. This dynamic is ultimately very similar to how translators who speak a rare language may demand higher wages.

**Empowering irreplaceable data labor:** Researchers and activists may be interested in leveraging the irreplaceability of certain types of data labor to reduce the power differential between these data laborers and technology companies. This may be achieved by recruiting expert users and niche groups in data strikes against data-driven technologies, such as asking those who are fans of certain movie genres to remove their ratings from recommender systems [84]. In situations in which technology companies wish to capture the totality of data generated by the public, members of groups currently unrepresented in existing datasets will have more power. It is important to note that the operative decision as to whether data labor from underrepresented groups has above-average or below-average replaceability depends on how technology companies plan to evaluate and deploy their technology (i.e. the choice of "training set").

**Empowering replaceable data labor:** For data labor that is easily replaceable, gaining and exerting labor power will be harder, and in particular will require larger group sizes [84]. Researchers and activists may look for ways that help users become irreplaceable, such as identifying and developing unique skills and knowledge as recommended by crowdwork researchers [41]. Additionally, policy-makers should take into consideration the role of societal inequities that prevent data producers from becoming irreplaceable, such as lack of means in gaining digital literacy for certain groups of the public[24]. In other cases, the most efficacious approach may be scaffolding collective action amongst a large pool of people who perform replaceable data labor. For instance, anyone who clicks search engine links is part of the massive pool of search engine trainers. This could be turned into an advantage by building global solidarity around this type of data labor, i.e. in pursuit of a "general data strike" [84].

## 3.7 Livelihood Overlap: Is Data Labor Part of Occupational Activities?

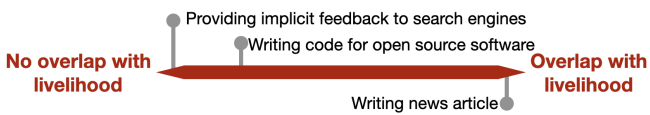

**Figure 6: The dimension of livelihood overlap: is data labor part of occupational activities? The positioning of data labor instances on the spectrum is approximate and may vary depending on the specific case.**

With the enormous popularity of foundation models like Chat-GPT and Stable Diffusion, this dimension has become quite salient: Does data labor overlap with data producers' occupational activities? Or put differently, do data-producing activities also help someone make a living? Providing implicit feedback to search engines and writing Wikipedia articles are rarely core occupational activities so they fall closer to the left end of the spectrum, i.e. no overlap with livelihood (Fig. 6.). On the right end of this spectrum, examples of data labor include producing images and artwork used in systems like Stable Diffusion and writing code that powers systems like GitHub Copilot. Both of these activities are usually undertaken as a core part of the data producers' occupational activities.

The technology companies that rely on data labor do not always have an employment relationship with data producers. For example, OpenAI scraped data from billions of webpages including news articles written by professional journalists who are hired by news organizations instead.

*3.7.1 Relationship with Power and Its Implications for Data Labor.* This dimension's relationship with power is complicated by the fact that technology organizations are able to collect the outcome of data labor without establishing formal employment relationships with data producers. While labor laws may provide data producers who perform data labor as a part of their job activities with legal venues to control the outcome of their work, these potential legal venues do not apply to data producers who are not employed by technology organizations.

**Empowering data labor with no overlap with livelihood**: It is important to note that people seeking to empower data labor should not ignore activities that do not overlap with one's livelihood. This type of data labor may create new professions as AI and other technologies need more and more accurate, high-quality datasets. For example, crowdwork has emerged from companies' needs for high-quality labels and datasets because hashtags and comments– an outcome of data labor activities with no overlap with livelihood– are no longer sufficient to advance model performance [32]. As such, researchers and activists may want to monitor emergent data labor opportunities that take non-traditional career forms and proactively identify opportunities to protect laborers' rights, such as fair wage and supportive working conditions.

**Empowering data labor with overlap with livelihood**: In general, when data labor has a high overlap with someone's livelihood, the actions of AI operating companies may pose a threat to that livelihood. The importance of this dimension for activists, policymakers, and researchers is straightforward: it will often be advisable to prioritize empowerment efforts where entire careers are disrupted by new systems. This would include, for instance, demanding models to be trained on artwork with artists' consent and developing tools that protect artists' styles from model training [74, 82]. Similarly, there is a great need for new technologies that will allow data producers to earn meaningful revenue from institutions like OpenAI if their work creates economic value through models like ChatGPT. For example, Shutterstock, a company that plans to sell Dall-E-generated images, announced its plan to compensate artists, in response to the artist community's criticism of the model's unapproved reuse of publicly available artwork [80]. While the details of the compensation mechanism have remained unclear, this example shows that the time may be ripe for data laborers' to pursue "back pay" from companies that monetize and benefit from their data labor.

However, compensation may not always be feasible or desirable in other contexts. The monetary value of a data point for the development of machine learning models is largely nebulous and speculative [16]. In these contexts, data laborers may be interested in pursuing non-monetary compensation, such as a voice to shape automation systems toward augmenting human labor instead of simply replacing it [12].

## 4 DISCUSSION

The six dimensions we identified and articulated above are only a starting point for understanding the rich variety of data labor activities. Our goal is to draw on existing knowledge to map out opportunities to organize and empower different kinds of data labor. The dimensions above suggest there may be low-hanging fruit to empower data labor, for instance where small design changes can make data labor more legible or provide stronger data control. Conversely, the dimensions above suggest certain data laborers may find it more difficult to gain power over technology companies and would need regulatory interventions, such as those perform open data labor.

While we have painted a broad picture of data labor, many, if not most, high-profile computing systems are complex and opaque to the public. As such, there may exist other dimensions that characterize data labor from technology companies' perspective. Below we discuss two such data labor characteristics. Additionally, we discuss the limitations of our synthesis.

### 4.1 Data Labor from a Technology Company Perspective

We discuss additional considerations regarding data labor that arise from taking the perspective of data-dependent technology operators. These factors are important to explore in future work, but compared to the core dimensions above are currently challenging to study because they will require comprehensive knowledge about how data is being collected, processed, managed, and used behind closed doors.

#### 4.1.1 Revenue Generation.

Quantifying data labor's relationship with revenue will assist technology operators in assessing to what extent their business is being subsidized by data producers and, thereby, gaining a more accurate quantitative understanding of their finances. Recent research in this area has made great strides in examining specific revenue streams of data; researchers have assessed Wikipedia's value to search engines [50], online communities [85], and commercial websites [61]. However, many other instances of data labor remain under-investigated due to the opacity and complexity of the ways data labor generates revenue for companies. Moreover, the particular causal link between data labor and firm revenue will vary depending on companies' business models. Future efforts that seek to comprehensively quantify data labor's value for technology companies must account for data's various connections to revenue generation, from ad sales to model performance improvement.

#### 4.1.2 Shelf Life.

Organizations that capture data labor may have specific requirements for how frequently new data labor must be captured, i.e. the 'shelf life' of data labor. For instance, systems that model real-time variables (e.g. misinformation classification and traffic estimation) need to collect data constantly, whereas some computer vision models may be able to use training data collected from years ago (e.g. ImageNet).

When data labor outputs do not have an expiration date, this is likely to reduce the leverage of data laborers, because it is very difficult in practice to delete expired data. To empower data labor, it may be fruitful for policies to limit how long certain records can be used, so data laborers can gain some leverage over technology operators over time. For example, GDPR has mandated that technology operators minimize the time period for which personal data is stored [1].

### 4.2 Limitations

As mentioned above, the list of dimensions we outlined in this paper is by no means exhaustive. As technology companies, practitioners, and researchers continuously identify new ways of using data, the characteristics of data labor will change. The fast-changing landscape of data labor highlights the urgency of identifying ways to include data laborers' voice in the design, development, and benefits of technology. Future work may expand our work and identify more dimensions and pathways toward this goal.

Another limitation of our synthesis is that we have largely focused on data labor for prominent technology organizations. However, data labor can generate revenue through improving a technology for other types of entities and organizations such as nonprofits, governmental agencies, and research communities. Such data labor, despite not being directly linked to large-scale capital generation for firms, may still hold societal importance (e.g. census records). Future work may characterize such instances of data labor to advance our taxonomy.

## 5 CONCLUSION

Through a synthesis of the literature on various types of labor and data, we introduce a roadmap for empowering data producers by constructing a formal definition and six core dimensions of data labor. For each dimension, drawing on the frameworks of empowerment and data leverage, we prescribe pathways for researchers, activists, and policymakers to empower the millions of people who generate data for prominent technology companies but currently have no say in how the data is used or who it benefits.

## ACKNOWLEDGMENTS

The authors would like to thank Isaac Johnson, Estelle Smith, Zach Levonian, and Leah Ajmani for providing feedback on an earlier version of this paper. This work was supported by Northwestern University's doctoral fellowship, NSF grants IIS-1815507 and IIS-1707296.